\documentclass[preprint2]{aastex}

\usepackage{graphicx,rotate}
\usepackage{natbib,lscape,longtable}
\usepackage{latexsym,amssymb,verbatim}

\shorttitle{Extragalactic D-species} \shortauthors{Bayet et al.}

\begin{document}


\title{Deuterated species in extragalactic star-forming regions}


\author{E. Bayet\altaffilmark{1}, Z. Awad\altaffilmark{1}
and S. Viti\altaffilmark{1}}

\email{eb@star.ucl.ac.uk}


\altaffiltext{1}{Department of Physics and Astronomy, University
College London, Gower Street, London WC1E 6BT, UK.}


\begin{abstract}

We present a theoretical study of the deuterated species
detectability in various types of extragalactic star-forming
regions based on our predictions of chemical abundances. This work
is motivated by the past and current attempts at observing
deuterated species in external galaxies such as NGC~253, IC~342
and the LMC. Here, we investigate the influence of the density,
the temperature, the FUV radiation field, the cosmic ray
ionisation, and the metallicity on the fractional abundances and
D/H abundance ratios of about 20 deuterated species. Without
modelling any particular source, we determined how the deuterium
chemistry behaves in different physical environments such as
starburst, cosmic-rays enhanced environments, low metallicity and
high redshift galaxies. In general, our predicted column densities
seem in good agreement with those derived from the current limited
dataset of observations in external galaxies. We provide, for the
first time, a list of key deuterated species whose abundances are
high enough to be possibly detectable by the Atacama Large
Millimeter Array (ALMA) and Herschel, as a function of galactic
nuclear activity and redshift.
\end{abstract}






\section{Introduction}\label{sec:intro}

The past few years have seen a boom in studies of molecular
deuteration in galactic low-mass and high-mass star forming
regions, triggered by the discovery of a large fraction of mono-,
doubly- and even multiply-deuterated molecules (e.g.
\citealt{Lore85, Vrti85, Walm87, Maue88, Jacq90, Geri92, Schi92,
Jacq93, VanDis95, Cecc98, Star99, Roue00, Pari02, Bacm03, Case03,
Vast04, Roue05, Miett09}). Studying deuterium chemistry,
especially in highly embedded regions, is important because in
such environments, CO is depleted and deuterated species are thus
driving the chemistry. In embedded regions in our Galaxy, a large
deuterium fractionation is observed, well above the elemental
abundance ratio D/H of $1.5 \times 10^{-5}$ \citep{Oliv03}. High
molecular deuteration can occur by two different mechanisms. One
involves gas phase chemistry and ion-molecule deuterium exchange
reactions taking place at low temperatures \citep{Wats80}. The
second mechanism is based on grain chemistry \citep{Tiel83}.

In this work, we investigate deuteration in extragalactic
star-forming environments, where the physical conditions can be
drastically different from what we see in our own Galaxy. To date,
very few detections of deuterated species have been attempted in
extragalactic environments (e.g. \citealt{Maue95, Mart06a}). In a
previous paper, \citet{Baye08a} investigated the non-deuterated
chemistry in a variety of extragalactic environments and obtained
theoretical predictions for the warmer and dense gas, traced by
non-deuterated species. Here we implement the mono-deuterated
species (hereafter called D-species) network to our existing one
used in \citet{Baye08a}, in order to investigate the differences
in the D-fractionation and the D/H ratios in regions of
star-formation (i.e. gas with n(H$_{2}$)$\geq 10^{5}$cm$^{-3}$) in
various types of galaxies (i.e starburst, cosmic-rays enhanced
environments, low metallicity and high redshift). We also aim at
guiding observers with potentially detectable deuterated tracers
of so called dense star-forming molecular gas in external galaxies
when using Herschel or the forthcoming ALMA, as based on the
strength of their chemical abundances.

The paper is divided in five sections; Section \ref{sec:model}
describes the model used and presents the choice of model
parameters used for mimicking various extragalactic environments.
Section \ref{sec:resu} presents the D-species fractional
abundances and D/H abundance ratios obtained in the various gas
components and environments investigated. Finally, in Section
\ref{sec:discu}, we discuss the results and conclude.

\section{Model descriptions}\label{sec:model}

We have used the same chemical model as the one used in
\citet{Baye08a}, but with an extended chemical network to include
mono-deuterated species. The chemical code, UCL\_Chem, is
presented in \citet{Viti99} and \citet{Viti04}. We refer the
reader to these papers for a full description of the code.

To summarise, UCL\_Chem model is a time and depth dependent
chemical model. As in \citet{Baye08a}, we first model the collapse
of a 10K core (Phase I); we then follow the chemical evolution of
the region once the stars are born (Phase II). The temperature in
both Phases is the typical average for starless and hot cores,
respectively, since here we are considering an ensemble of cores
within a large region of space. \citet{Baye10b} do in fact explore
in great detail the influence of changes in temperature and
density within a single hot core when subject to various
environmental conditions. They find that thermal variations due to
temperature structure would flatten out when averaged over a large
number of cores.

The presence of an infrared source in the centre or in the
vicinity of the region is simulated by subjecting the gas and the
dust to an increase in temperature. In our models, we have assumed
equal dust and gas temperatures. This is justified since in Phase
II the opacity is very high (see values listed in Table
\ref{tab:4}) and thus the thermal coupling between gas and dust is
valid.

In both phases, the chemical network, now improved with D-species,
is based on more than 4000 chemical reactions adapted from the
UMIST 2006 database \citep{Mill97,LeTe00,Wood07} involving about
300 species (92 of them being deuterated) of which 62 are surface
mantle species. The relevant surface reactions included in this
work involve mainly simple hydrogenation, apart from the formation
of the CH$_{3}$CN which is believed to be formed via reactions
involving HCN on grains \citep{Garr06}. We note that our approach
for surface reactions is rather simplified with respect to that
adopted by \citet{Garr06} and \citet{Caza10}. Nevertheless, a
qualitative comparison of the mantle composition at the end of
Phase I with the results from \citet{Caza10} mantle show good
agreement (c.f. H$_{2}$O and HDO abundances on grains as a
function of the radiation field for example). As a first
approximation and to keep the chemical network to a reasonable
size, we include only the mono-D species. D$_{2}$CO and its
corresponding ion are the only doubly-D-species in our network.

The deuterium chemical network is similar to that of
\citet{Robe00a, Robe00b} and \citet{Robe04}. We assume that D
atoms react in the same way as H atoms when reacting with other
atoms and molecules.

One of the outputs of the UCL\_Chem is the fractional abundance
(with respect to the total number of hydrogen nuclei and as a
function of time) of gas and surface species from which we
calculate the D/H ratios.

The above assumptions (i.e. no doubly-D-species except D$_{2}$CO
formed on grain nor triply-D-species included in the code) do
affect the reliability of the predictions for a couple of species
namely D$_{2}$CO, H$_{2}$D$^{+}$ and N$_{2}$D$^{+}$ (see
\citealt{Robe03}). On the other hand, only mono-D-species have
been observed so far in extragalactic environments, and the
relative doubly-D/mono-D or triply-D/mono-D species abundance
ratios seen in galactic sources are small (see e.g.
CD$_{3}$OH/CH$_{2}$DOH = 0.46 and CHD$_{2}$OH/CH$_{2}$DOH = 0.2
from \citealt{Pari04}, ND$_{2}$H/NH$_{2}$D = 0.05 from
\citealt{Roue05}, and, D$_{2}$CO/HDCO=0.02-0.33 from
\citealt{Roue03}).

To investigate the deuterium chemistry in high-mass star-forming
regions in various extragalactic environments, we have run a grid
of 15 chemical models (see Tables \ref{tab:1}, \ref{tab:3} and
\ref{tab:2}). For each type of galaxy (see below), we have studied
the deuterium chemistry coming from a moderately dense
($n(H_{2})=10^{5}$cm$^{-3}$ - index 1 used in tables and figures),
a dense ($n(H_{2})=10^{6}$cm$^{-3}$ - index 2) and a very dense
gas ($n(H_{2})=10^{7}$cm$^{-3}$ - index 3) component. This has
allowed us to potentially distinguish which gas component produces
the highest fractional abundances of D-molecules, in each type of
environment. This also allowed us to determine which gas component
is likely to be the best match to the observed extragalactic D/H
ratios (see Sect.\ref{sec:resu}).

For reproducing the various types of galaxies, we have assumed the
following :
\begin{itemize}
\item [] The \emph{\textbf{Normal Spiral}} (NS) case assumes
standard parameter values (symbol ``ST'') similarly as previously
performed by Bayet et al. (2008; see their Tables \ref{tab:1} and
\ref{tab:3}). \item [] For the \emph{\textbf{Starburst}} (SB)
case, we have used a FUV radiation field of 1000 times the
standard value, and a temperature of 500 K, in Phase II, instead
of 300 K. These models aim at mimicking environments such as
NGC~253 or M~82. \item [] The \emph{\textbf{Cosmic-ray enhanced
environments}} (SB$+$) case is represented in our models by a
cosmic ray ionisation rate of 100 times the standard value. This
type of models aims at reproducing supernovae and massive star
formation as seen in galaxies such as Arp 220 (supernovae rate of
4 year$^{-1}$, see e.g. \citealt{Parr07}). \item [] The
\emph{\textbf{Low metallicity}} (Low-met) case is represented, in
our models, by a decrease in the metallicity to a fifth of the
solar value coupled with an increase of the gas-to-dust mass ratio
by a factor of 5. Indeed, in the code, quantities such as
metallicity, optical depth ($A_{v}$) and the H$_{2}$ formation
rate coefficient are coupled (see values presented in Table
\ref{tab:4} for each model). In this galaxy type, we have also
assumed an increase of FUV radiation field of 1000 times the
standard value, and a temperature of 500 K since we are
particularly interested in sources such as IC~10, which also host
a strong star-formation activity. \item [] Finally, the
\emph{\textbf{High redshift}} (High-z) case is represented by
models with a metallicity (and its coupled quantities) of a fifth
solar as measured in IC10 \citep{Zari94}, which is a local dwarf
galaxy often used as a good template for archetypical higher-z
galaxy populations (e.g. \citealt{Madd97, Lero06, Yin10}); an
increase of FUV radiation field by 1000 times the standard value;
a temperature of 500 K and an increase of the cosmic ray
ionisation rate by a factor of 100 times the standard value.
Indeed, in distant objects (at redshifts greater than 0.1), it is
expected to often have a combination of several nuclear activities
such as AGN, SB and supernovae (see for instance \citealt{Seym09}
or the studies on the quasar APM08+279 located at z$\sim4$ from
\citealt{Weis07} and \citealt{Riec09}). We note that we may indeed
underestimate the metallicity at high redshift in our study since
optical observations of quasars at z$\sim 6$ (e.g.
\citealt{Jian07, Juar09}) recently revealed that a solar
metallicity value is more likely than a sub-solar one for
reproducing their observations. However, we emphasize that we have
not tried in our present models to match the physical conditions
for any high-z source in particular but that we model here what
one can consider as an archetypical high redshift environment.
Since we know that the early Universe is more populated by dwarf
galaxies than any other galaxy type (e.g. \citealt{Whit91}), we
adopt the values for IC10 as our high redshift standard.
\end{itemize}

All the details of the parameters used in the 15 models are
summarized in Tables \ref{tab:1} to \ref{tab:4}.

\section{Sensitivity of deuterated chemical abundances and
D/H abundance ratios to variations in the physical and chemical
parameters.}\label{sec:resu}

The main aim of this work is to study the sensitivity of
deuterated species in models of star-forming regions to variations
in physical and chemical parameters that may be characteristics of
various galaxy types. A collection of \textbf{likely} detectable
tracers as seen by the strength of their chemical abundances will
be given in Subsect. \ref{subsec:pred}, but first we simply
analyse the trends the deuterium chemistry follows as the
environment (i.e. galaxy type) and the gas density vary. These
trends are summarized in Figs. \ref{fig:1} to \ref{fig:6}.

\subsection{Deuterated fractional abundances sensitivity}

\subsubsection{Influence of the gas density}

\begin{figure*}
    \centering
    \includegraphics[width=15cm]{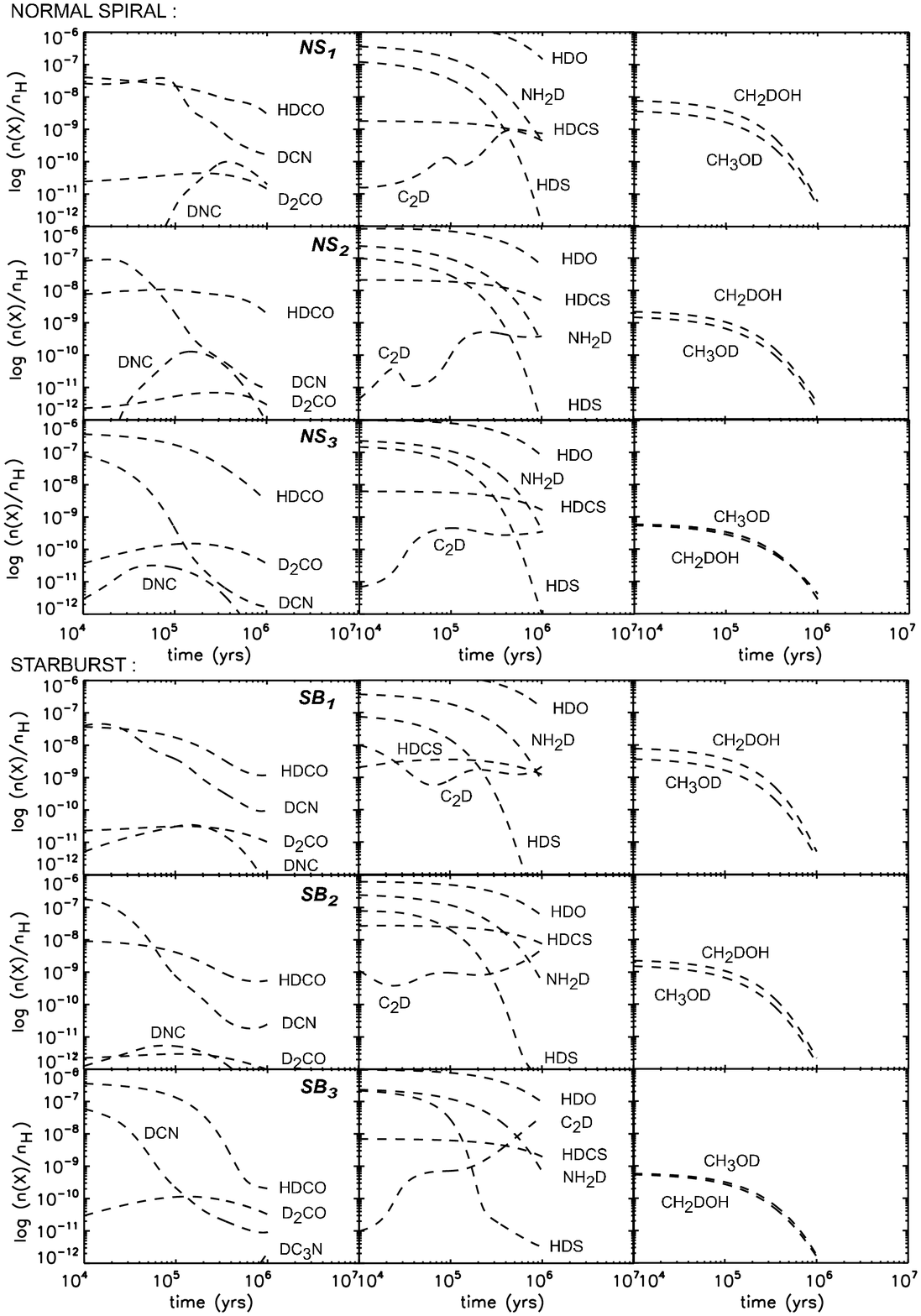}
    \caption{Fractional abundances as a function of time of various D-species for
    the normal spiral case (top three plots) and the
    starburst case (bottom three plots). For each case,
    we have plotted the three gas components seen in Table
    \ref{tab:2} i.e the
    lowest density gas component (n(H$_{2}$)=$10^{5}$cm$^{-3}$, top),
    n(H$_{2}$)=$10^{6}$cm$^{-3}$ (middle) and the
    densest gas component (n(H$_{2}$)=$10^{7}$cm$^{-3}$, bottom).
    Fractional abundances are expressed with respect to the
    total number of hydrogen nuclei n$_{\rm H}$ and the limit
    of detectability has been taken to be [n(X)/n$_{H}$] =
    $1\times 10^{-12}$.}\label{fig:1}
\end{figure*}

\begin{figure*}
    \centering
    \includegraphics[width=15cm]{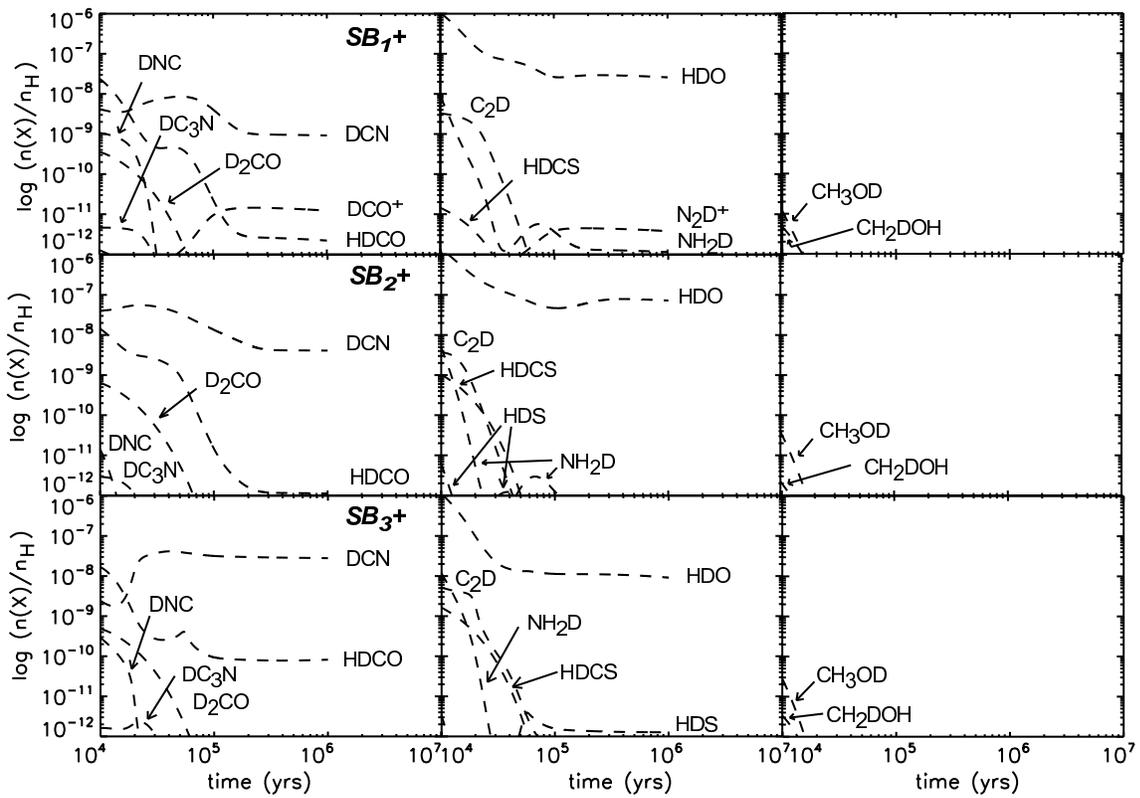}
    \caption{Fractional abundances as a function of time of various D-species for
    the cosmic-rays enhanced case (see caption of
    Fig. \ref{fig:1}).}\label{fig:2}
\end{figure*}

\begin{figure*}
    \centering
    \includegraphics[width=15cm]{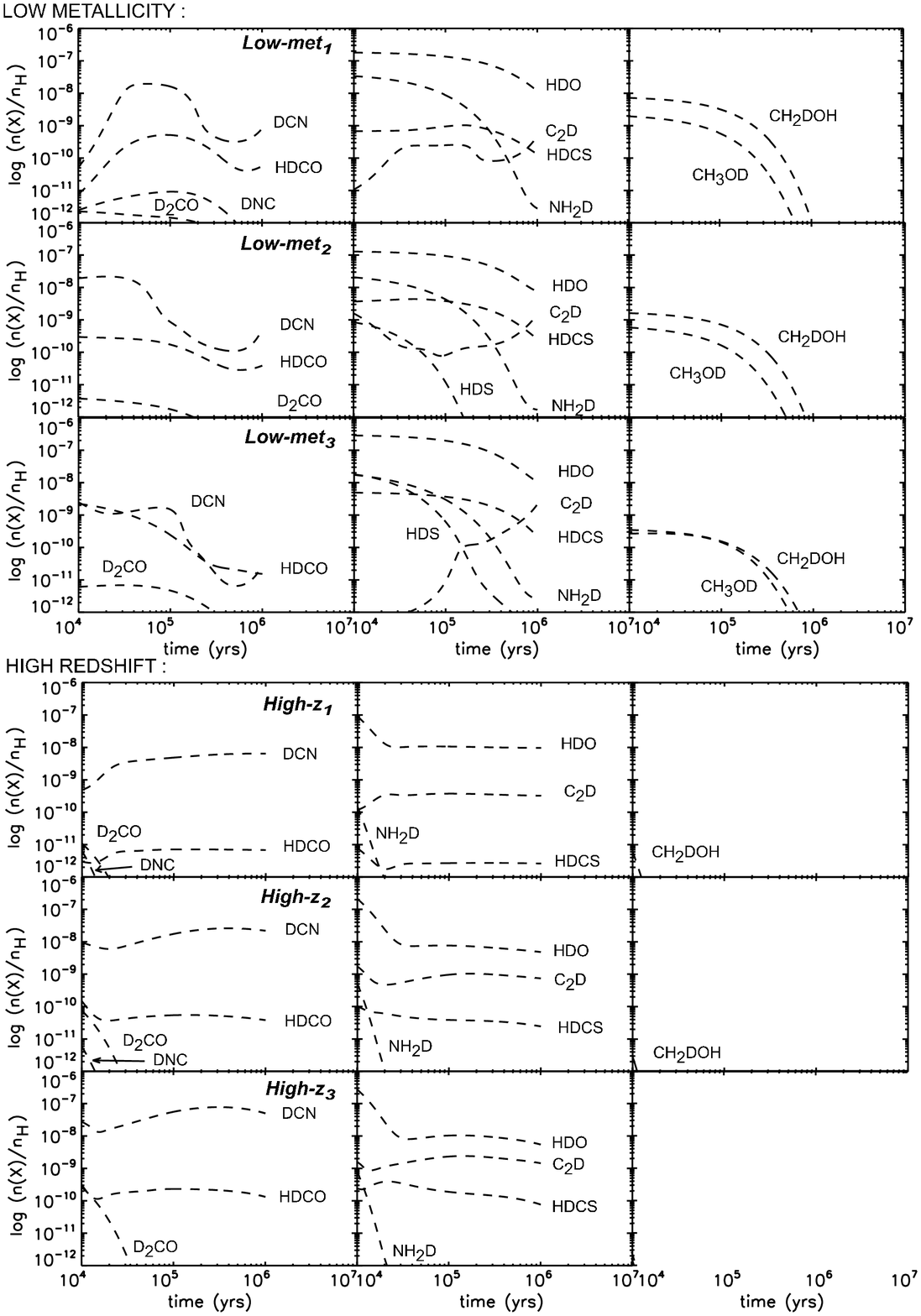}
    \caption{Fractional abundances as a function of time of various D-species for
    the low metallicity case (top three plots) and the
    high redshift case (bottom three plots). See the caption of
    Fig. \ref{fig:1}.}\label{fig:3}
\end{figure*}

When the gas density increases from n(H$_{2})=10^{5}$cm$^{-3}$ to
n(H$_{2})=10^{7}$cm$^{-3}$, regardless of the galaxy type, we can
divide deuterated species into three categories: those which are
insensitive to density changes (e.g. HDO, NH$_{2}$D and HDCS),
those that increase with density (e.g. HDS, C$_{2}$D and
CH$_{3}$OD) and those which show a decrease in their fractional
abundances (e.g. DCN, DNC and CH$_{2}$DOH). The fractional
abundances of the D-species belonging to the first category
actually tend to converge after $\sim 10^{5}$yrs, making these
species good D-tracers of dense gas (n(H$_{2})\geq
10^{5}$cm$^{-3}$) whatever the environments since they show
abundances above 10$^{10}$. On the contrary, we show in Figs.
\ref{fig:1} to \ref{fig:3} that especially DCN and DNC decrease
with an increase of density (e.g. compare Models NS$_{1}$ and
NS$_{3}$). This is no surprise since the same behavior is seen for
both HCN and HNC when the density increases \citep{Awad10}. This
phenomenon reinforces the conclusions of \citet{Baye09c} that HCN,
HNC and their deuterated counterparts are not particularly good
tracers of the very dense gas in galaxies. HDS, C$_{2}$D and
CH$_{3}$OD increase with density. The enhancements in their
fractional abundance, by more than a factor of 5 (Figs.
\ref{fig:1} to \ref{fig:3}) are particularly interesting since
observing their molecular emissions may allow us to discriminate
between the different gas components in a galaxy.

\subsubsection{Influence of the FUV radiation field and the
temperature (SB case)}

Variation in FUV or temperature do not lead to significant changes
in abundances for species such as HDO, NH$_{2}$D, DCN, D$_{2}$CO,
HDCS, CH$_{2}$DOH and CH$_{3}$OD (i.e. variations in chemical
abundances less than a factor of 2-3). However, HDCO and HDS
abundances decrease between a factor of 5 and 15 when the FUV
radiation field increases. This is due to dissociations by FUV
photons. At early times ($\leq 10^{5}$ yrs), and at low density
(n(H$_{2})= 10^{5}$cm$^{-3}$) the abundance of C$_{2}$D increases
by three orders of magnitude with the increase of both FUV
radiation field and temperature. Hence, this species is likely to
be a good D-tracer of excitation in galaxies since very sensitive
to changes in both temperature and FUV radiation field. At later
times, the increase of the C$_{2}$D abundance is less pronounced,
but still significant (i.e. about two orders of magnitude). On the
other hand, DNC shows the largest decrease i.e. by more than one
order of magnitude, when the FUV radiation field and the
temperature are increased (linked as seen in Table \ref{tab:3}).
This may be due to cosmic-ray-induced photodissociation, which has
a significant temperature dependence.

\subsubsection{Influence of the cosmic ray ionisation rate (SB$+$ case)}

Increasing the cosmic ray ionisation rate ($\zeta$) by a factor of
100 (Models SB$_{1}+$ to SB$_{3}+$ - See Table \ref{tab:2} and
Fig. \ref{fig:2}) leads to most abundances been significantly
reduced by several orders of magnitude (e.g. HDCS, NH$_{2}$D,
D$_{2}$CO, HDCO, CH$_{2}$DOH, HDO, DNC). As already found by
\citet{Baye08a}, by \citet{Meij05b}, and \citet{Meij06}, the
effects of increasing the cosmic ray ionisation rate are complex,
and they generally lead to a chemistry that approaches
steady-state more quickly (see Fig.\ref{fig:2}). The general
decrease of abundances seen when $\zeta$ increases is probably due
to a release of reactive ions such as ionized carbon, reacting
quickly with many oxygen-bearing species such as HDCO, D$_{2}$CO,
etc. However, some species such as DCO$^{+}$, N$_{2}$D$^{+}$ (both
at low density only), DC$_{3}$N and DCN (whatever the gas
component) show abundances increased by several orders of
magnitude with $\zeta$. In the case of nitrogen-bearing species,
we suspect that the main reservoir of nitrogen, N$_{2}$, becomes
ionized and gives rise to N$^{+}$, a reactive species that
promotes the formation of nitrogen-bearing species. This behaviour
is also found for the hydrogen counterparts (e.g.
\citealt{Baye08a}). What is interesting here is that D-species
seem to be affected in the same way as their hydrogen counterpart.
Species such as N$_{2}$D$^{+}$ and DCN are thus likely to be good
D-tracers of cosmic ray ionisation-excited environments since they
show high fractional abundances and are sensitive to their change.
In principle, we thus can use deuteration to constrain $\zeta$
although the effects of high cosmic rays ionisation rates on the
chemistry are very complex (Bayet et al. 2010, in prep) and highly
dependent on initial conditions as well as the available gas
coolants \citep{Baye10a}.

\subsubsection{Influence of the metallicity (Low-met case)}

Low metallicity environments such as those described in Sect.
\ref{sec:model} (see Models Low-met$_{1}$ to Low-met$_{3}$ in
Table \ref{tab:2}), lead to few surprises for most species in that
their abundance is simply reduced accordingly (see Fig.
\ref{fig:3} for HDCO, D$_{2}$CO, DNC, NH$_{2}$D and HDS). However,
some species, such as C$_{2}$D or DCN, show an increase in their
abundances with a decrease of the metallicity whereas some other
species such as HDCS, CH$_{2}$DOH, CH$_{3}$OD do not seem
sensitive to metallicity changes. Therefore these latest species
(potentially) are good D-tracers of dense gas (since they have
high abundances) whatever the metallicity (especially interesting
at high redshift), similarly to their hydrogen counterpart (such
as CH$_{3}$OH) already noted by \citet{Baye08a} and
\citet{Roel07}.

\subsubsection{Influence of the above parameters coupled together (High-z case)}

In extreme conditions such as those used for mimicking high
redshift environments (see Models High-z$_{1}$ to High-z$_{3}$ in
Table \ref{tab:2} and Fig. \ref{fig:3}), most of the chemical
abundances of D-species decrease: CH$_{3}$OD, HDS disappear
completely from Fig. \ref{fig:3} as compared to the normal spiral
case (see Fig. \ref{fig:1}); NH$_{2}$D, D$_{2}$CO, DNC, and
CH$_{2}$DOH have chemical abundances above the limit of
detectability only at early time (i.e. before 1.5-3$\times 10^{4}$
yrs) then they drop severely at later time; HDCS, HDO, HDCO, even
if still above the limit of detectability at all times, decrease
by more than two orders of magnitude in high redshift environments
as compared to the normal spiral case. These species nonetheless
remain the best D-tracers of dense gas components in high redshift
environments since very sensitive to the parameters changes. Only
DCN and C$_{2}$D show fractional abundance either boosted by up to
four orders of magnitude as compared to the Models
NS$_{1}$-NS$_{3}$ or stay unchanged, respectively. In the case of
the high-z, the formation of DCN is dominated by the reaction D +
HCN which is inefficient in our models for normal spirals because
of the deficiency of atomic deuterium. In the models for normal
spirals, the fractional abundance of D is indeed five orders of
magnitude lower than in the models for high-z cores.

\subsection{D/H abundance ratios sensitivity}

\begin{figure*}
    \centering
    \includegraphics[width=11cm]{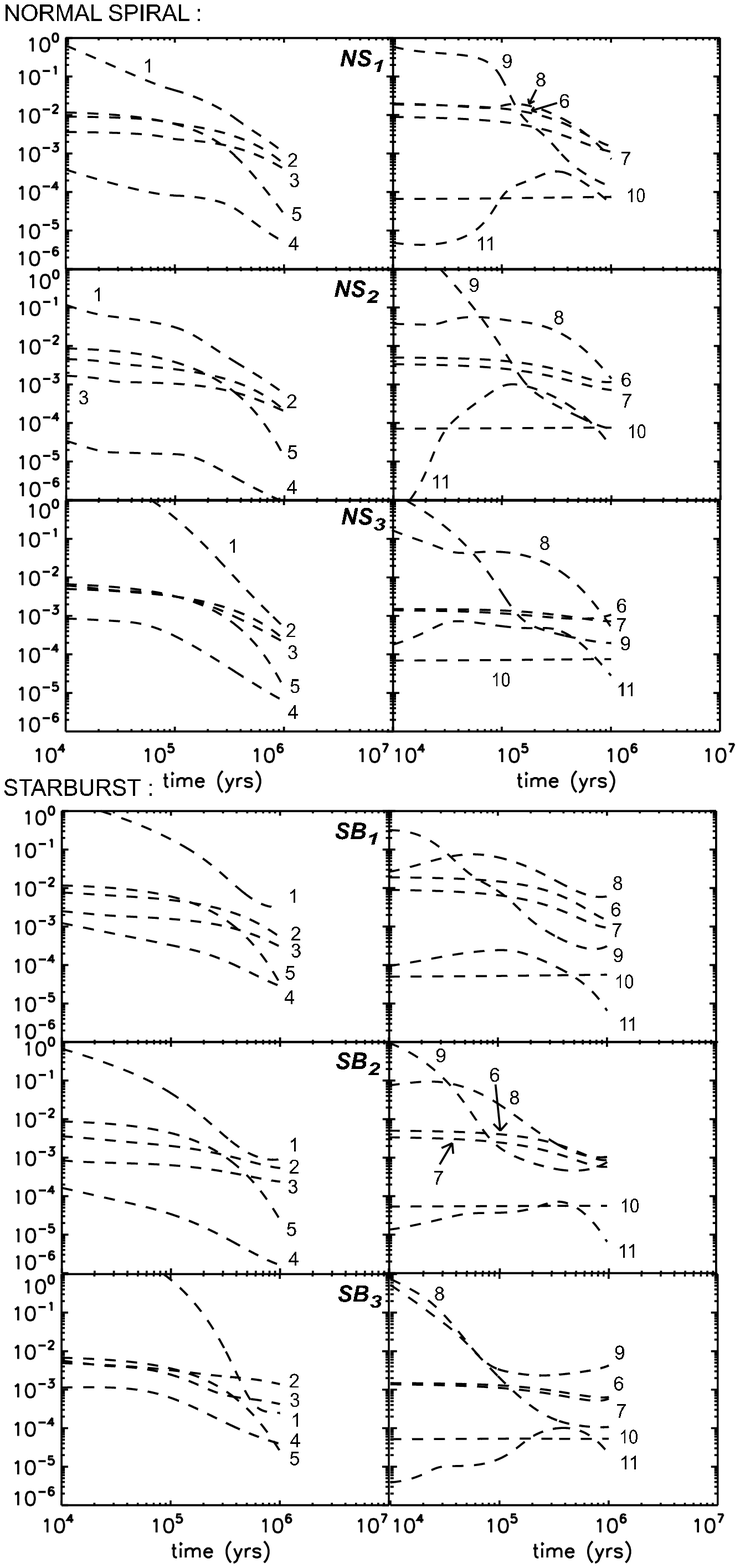}
    \caption{D/H abundance ratios as a function of time for
    various D-species in
    the normal spiral case (top three plots) and in the
    starburst case (bottom three plots). See the caption of
    Fig. \ref{fig:1}. 1: HDCO/H$_{2}$CO, 2: N$_{2}$D$^{+}$/N$_{2}$H$^{+}$,
    3: DCO$^{+}$/HCO$^{+}$, 4: D$_{2}$CO/H$_{2}$CO, 5: NH$_{2}$D/NH$_{3}$,
    6: CH$_{2}$DOH/CH$_{3}$OH, 7: CH$_{3}$OD/CH$_{3}$OH,
    8: CH$_{2}$DCN/CH$_{3}$CN, 9: DCN/HCN, 10: H$_{2}$D$^{+}$/H$_{3}^{+}$,
    11: DNC/HNC. }\label{fig:4}
\end{figure*}

\begin{figure*}
    \centering
    \includegraphics[width=11cm]{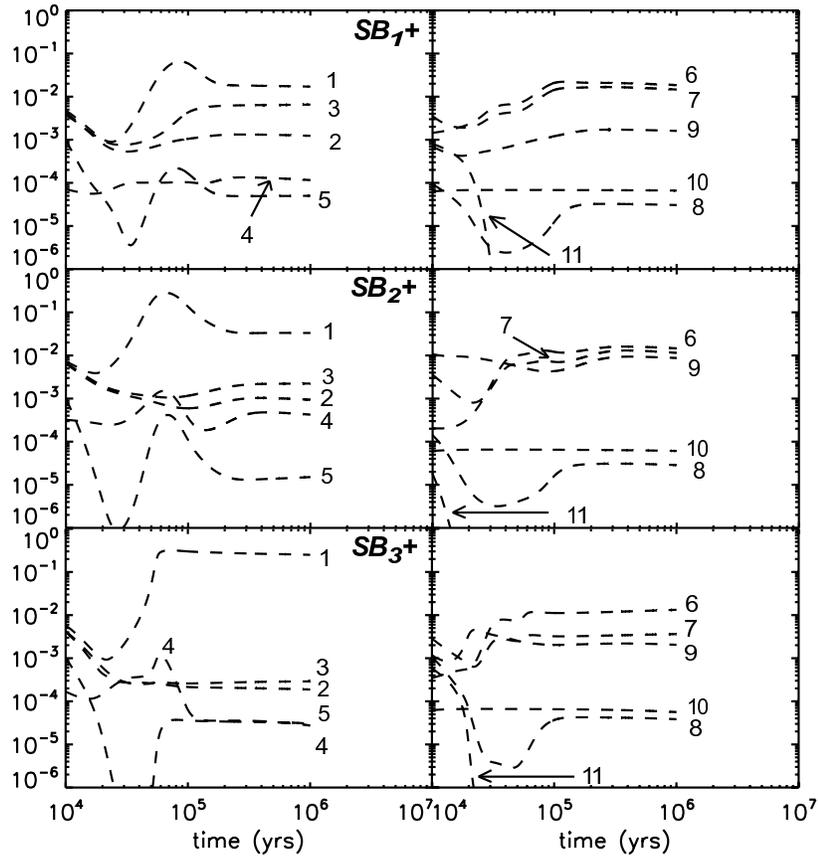}
    \caption{D/H abundance ratios as a function of time for various D-species in
    the cosmic-rays enhanced case (see captions of
    Figs. \ref{fig:1} and \ref{fig:4}).}\label{fig:5}
\end{figure*}

\begin{figure*}
    \centering
    \includegraphics[width=11cm]{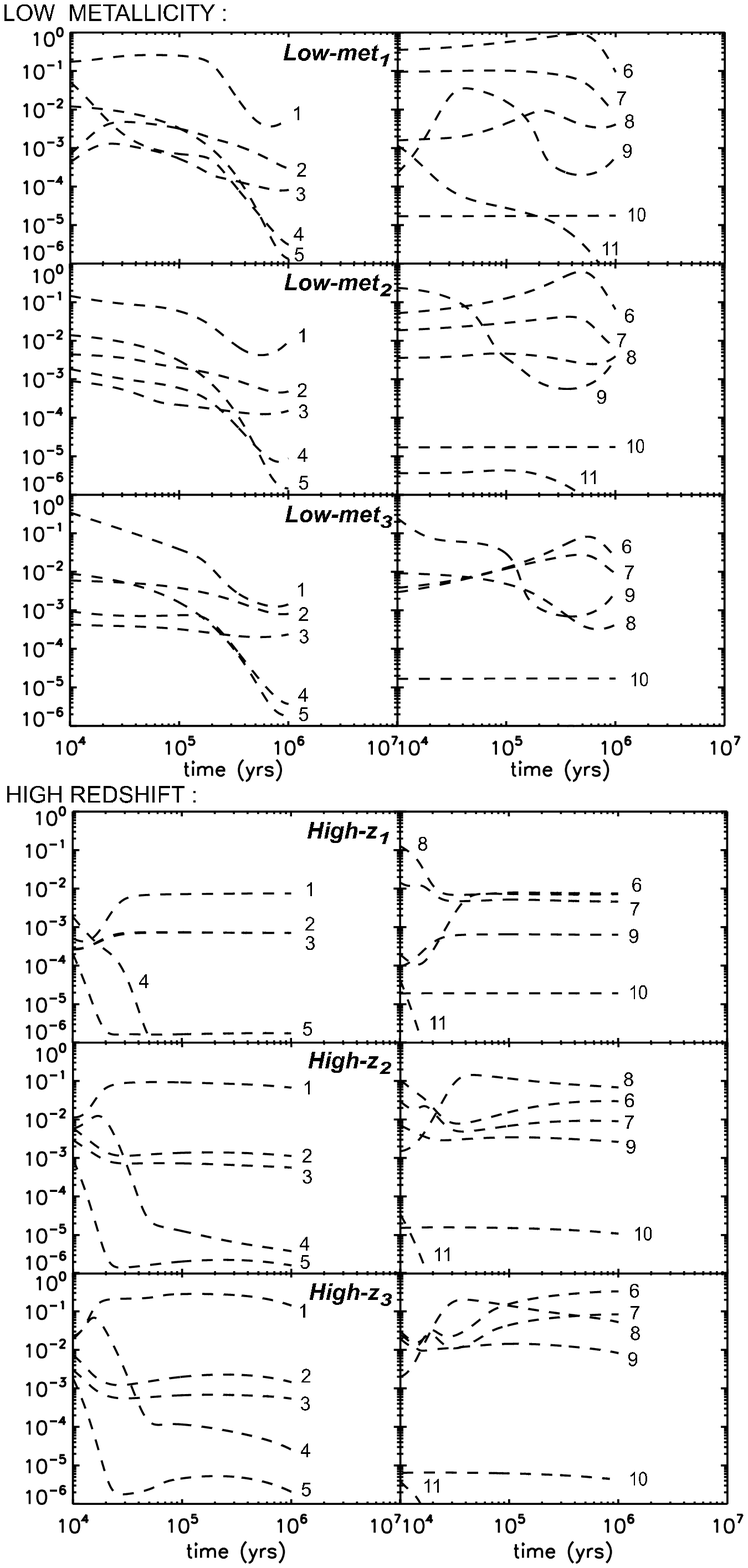}
    \caption{D/H abundance ratios as a function of time for various D-species in
    the low metallicity case (top three plots) and the
    high redshift case (bottom three plots). See the captions of
    Figs. \ref{fig:1} and \ref{fig:4}.}\label{fig:6}
\end{figure*}

As seen in Figs. \ref{fig:4} to \ref{fig:6}, the D-fractionation
calculated from the chemical abundances of all species do not show
similar D/H ratio values, regardless of the environment. In
protostellar galactic cores, such as IRAS 16293-2422, such
behaviour is already well-known (e.g. \citealt{VanDis95, Cecc98,
Pari04}). Authors indeed observed NH$_{2}$D/NH$_{3}$=0.1 whereas
they obtained DCO$^{+}$/HCO$^{+}$=0.009. They interpret the higher
D/H ratios they obtained as coming from a cooler and more extended
gas located in the envelope around the source and not coming from
the hot gas in the core. In our study, we can see (Fig.
\ref{fig:4}) that the highest ratios are indeed obtained, on
average, for the galactic cases (i.e. Models NS), for Model
NS$_{1}$ which has got the smallest density (see Table
\ref{tab:5}).

In roughly all the models, there are three groups of molecules
which converge to similar D/H values: the group including
DCO$^{+}$/HCO$^{+}$, N$_{2}$D$^{+}$/N$_{2}$H$^{+}$ and
HDCO/H$_{2}$CO; the group containing NH$_{2}$D/NH$_{3}$ and
D$_{2}$CO/H$_{2}$CO and the group containing
CH$_{2}$DOH/CH$_{3}$OD, CH$_{2}$DCN/CH$_{3}$CN,
CH$_{3}$OD/CH$_{3}$OH. D/H however varies by several orders of
magnitude amongst those groups. With the current limited D/H
dataset on galactic and extragalactic sources, confirmation of
such results is not possible.

The H$_{2}$D$^{+}$/H$_{3}$$^{+}$ does not vary as much as the
other D/H ratios from a galaxy type to another. On average the
largest discrepancy between D/H ratios group estimates is shown to
be for the SB$+$ case where the spread between all the D/H values
is the largest (see Fig.\ref{fig:5}).

Despite these uncertainties and for providing future observers
with D/H values, steady-state D/H ratios for various molecules are
listed in Table \ref{tab:5}.

\subsection{Predictions for observers}\label{subsec:pred}

In Table \ref{tab:5}, we list the D-species that should be likely
detectable in external galaxies as based on the values obtained
for their chemical abundance. We define the limit of detectability
as [n(X)/n$_{H}$] = $1\times 10^{-12}$, as is typical for dense
gas molecules in the Milky Way. Of course, detectability will also
depend on the excitation conditions as well as the geometry of the
source. According to the estimates of \citet{Lint05}, however
these abundances may be sufficiently large that unresolved active
galaxies should be likely detectable in these species even at high
redshift.

Of particular interest are HDO, DCN and HDCO which are very
abundant ([n(X)/n$_{H}$]$> 10^{-10}$) in all the investigated
environments. These species are ideal molecules to be used for a
first observational campaign aiming at validating the model
predictions described here.

A second category of interesting species includes DCO$^{+}$,
DC$_{3}$N, DNC and N$_{2}$D$^{+}$ as they seem to arise only from
one gas component in a single galaxy type: SB$_{3}+$ (for
DCO$^{+}$), SB$_{1}$ (for DC$_{3}$N), NS$_{3}$ (for DNC) and
SB$_{3}+$ (for N$_{2}$D$^{+}$). In a similar vein, C$_{2}$D and
HDCS are predicted to be excessively reactive to the presence of
cosmic rays ionisation whereas they survive easily in FUV-enhanced
environments. HDS is more sensitive to FUV changes than C$_{2}$D
and HDCS and it survives to both FUV and cosmic ray ionisation as
long as the gas is dense. Here, HDS may thus be a good
discriminant of the different gas components in SB and SB$+$
environments since it shows high abundances for the densest gas
component. However, as soon as the metallicity drops, it becomes
undetectable whereas HDCS and C$_{2}$D abundances are still high.

The third category of species contains D$_{2}$CO, CH$_{2}$DOH and
CH$_{3}$OD. These molecules do not survive in high-redshift nor in
low-metallicity galaxies. The fact that these species survive
nonetheless in SB sources may reflect the poor influence of FUV
photons on highly embedded material.

Species such as CH$_{2}$DCN and H$_{2}$D$^{+}$ are likely
undetectable, whatever the conditions since their chemical
abundances are low.

Finally, DCN, HDO, HDCO and to a lesser extent D$_{2}$CO,
DCO$^{+}$ and NH$_{2}$D seem also to be good candidates i.e.
abundant enough at various redshifts.

Of course, our predictions are qualitative in nature since based
only on an abundance criterion. For more quantitative predictions,
radiative transfer models, as well as the knowledge of the source
geometry, are needed. DCN, HDCO, D$_{2}$CO, DCO$^{+}$ and
NH$_{2}$D expected line intensities can not be derived due to the
lack of collisional rates. For HDO however, we have been able to
derive very rough estimates of line intensities using
RADEX\footnote{See http://www.sron.rug.nl/~vdtak/radex/radex.php}
developed by \citet{VanderTak07} and using the collisional rates
from the Leiden Atomic and Molecular Database (LAMDA\footnote{See
http://www.strw.leidenuniv.nl/~moldata/}). We have assumed a
plane-parallel geometry, used Model SB$_{1}$ HDO fractional
abundances for describing the local Universe abundances, and Model
High-z$_{1}$ HDO fractional abundances for the z$>$ 1 case. We
have converted the predicted HDO fractional abundances from both
models into column densities as described in Sect. \ref{sec:discu}
(see the footnote), using the A$_{\rm v}$ values listed in Table
\ref{tab:4}. The HDO column densities obtained have then been used
in RADEX, in addition to the gas density and the kinetic
temperature as listed in Table \ref{tab:3}. A FWHM of 50
kms$^{-1}$ has been assumed, as seen typically in extragalactic
molecular line measurements (e.g. \citealt{Maue95, Isra03,
Mart06a, Baye09c} and \citealt{Alad10}). For Models SB$_{1}$, we
have obtained values ranging from 1.8$\times 10^{-4}$ Kkms$^{-1}$
(for the HDO(5$_{3,2}$-6$_{1,5}$) line emitting at 356 GHz) up to
9.5$\times 10^{3}$ Kkms$^{-1}$ (for the HDO(2$_{0,2}$-1$_{1,1}$)
transition emitting at 490 GHz). Beam and distance dilution
correction factors have to be applied to these values. Following
\citet{Lint05} formula (see the Appendix), we have assumed a
distance of 3 Mpc (such as M 82) as well as a beam size of 12
arcsec as listed for ALMA-band 8\footnote{See primary beam size
values in
http://www.eso.org/sci/facilities/alma/observing/specifications/}
(i.e. at 490 GHz). With such assumptions, the
HDO(2$_{0,2}$-1$_{1,1}$) line intensity can be reproduced by $\sim
10^{5}-10^{8}$ cores, depending on the source size considered. If
we perform the same calculation but for the high redshift case
(i.e. HDO fractional abundances from Model High-z$_{1}$), for a
galaxy located at z=4.7 \citep{Cari02}, and increasing the ALMA
beam to 56 arcsec (since the frequency line drops to ALMA-band 1),
we thus need $\sim 10^{9}$ cores to reproduce the
HDO(1$_{0,1}$-0$_{0,0}$) line emitting at 465 GHz, which shows the
strongest RADEX line intensity estimate of 2.09$\times 10^{1}$
Kkms$^{-1}$. In this case, HDO may not, of course, be detectable.

\section{Discussion and conclusions}\label{sec:discu}

Although we do not aim at modelling any specific source, we have
qualitatively compared our models with extragalactic D-species
observations published so far. \citet{Maue95} obtained upper
limits for the emission of the DCN J=2-1 line toward NGC 253 and
IC 342. More recently, \citet{Mart06a} observed DCO$^{+}$, DCN,
DNC and N$_{2}$D$^{+}$ in NGC 253. When we convert these
observations and our predicted fractional abundances into
approximate column densities\footnote{We performed such conversion
by simply multiplying our predicted fractional abundance by the
column density of hydrogen at the relevant visual extinction.}, we
find a general good agreement between the two sets within a factor
of $\leq$ 2-3, even if one single model is not able to reproduce
consistently all the data. As already shown in \citet{Baye09c},
several gas components are needed to account for the molecular
emission observed in external galaxies. More precisely, in NGC 253
(data from both \citealt{Maue95} and \citealt{Mart06a}), the Model
SB$_{1}+$ best reproduces the observed DCO$^{+}$ column density
(less than a factor of 3). This indicates that a cosmic
ray-enhanced environment may be present in the nucleus of NGC 253.
Model NS$_{3}$ best reproduces the observed DCN column density in
NGC 253, within a factor of 1. Finally, all models in our grid,
except Model NS$_{1}$, are able to reproduce the observation of
DNC (N(DNC) $\leq 7.2 \times 10^{10}$ cm$^{-2}$). It is doubtful
whether we can compare the observed and the predicted column
densities for N$_2$D$^{+}$ because it is expected that
doubly-deuterated species, not included in our current chemical
network, might provide important formation and destruction
chemical routes for N$_2$D$^{+}$ \citep{Robe03}. With such
restriction in mind, it appears however that all the models except
Model SB$_{1}+$ can reproduce the observed upper limit of
N(N$_2$D$^{+}$) $\leq 9.4\times 10^{10}$ cm$^{-2}$ reported in
\citet{Mart06a}. For IC 342, only one D-species has been observed
(i.e. DCN) and Model NS$_{3}$ gives the closest predicted column
densities to the observed values (factor of 11). This factor is
higher than in the case of NGC~253 because the DCN detection
suffers from larger uncertainties in IC~342 (see \citealt{Maue95})
as compared to the detection obtained in NGC~253.

Due to the lack of additional observations in extragalactic
environments, we are not able to further refine our predictions.
Nevertheless, our simple approach offers some significant insights
on the major trends of the deuterium chemistry when subject to
various physical conditions. Our models can also potentially help
discriminate cold and dense star-forming gas between SB, SB$+$,
low-metallicity, etc activity. These predictions may be of
particular interest when looking at extragalactic very dense
high-mass star-forming regions where CO is expected to be depleted
onto grains (similarly to what it is seen in our own Galaxy), and
where thus D-species drive the chemistry.

\section*{Acknowledgments}

EB acknowledges financial support from STFC. ZA would like to
thank both the ORS and the Perren studentship schemes for funding.
The authors thank the referee for helpful comments on the original
version of this paper.

\bibliographystyle{apj}
\bibliography{references}

\begin{thebibliography}{66}
\expandafter\ifx\csname natexlab\endcsname\relax\def\natexlab#1{#1}\fi

\bibitem[{{Aladro} {et~al.}(2010){Aladro}, {Martin-Pintado}, {Martin},
  {Mauersberger}, \& {Bayet}}]{Alad10}
{Aladro}, R., {Martin-Pintado}, J., {Martin}, S., {Mauersberger}, R., \&
  {Bayet}, E. 2010, ArXiv e-prints 2010arXiv1009.1831A

\bibitem[{{Awad} {et~al.}(2010){Awad}, {Bayet}, \& {Viti}}]{Awad10}
{Awad}, Z., {Bayet}, E., \& {Viti}, S. 2010, \mnras, in prep.

\bibitem[{{Bacmann} {et~al.}(2003){Bacmann}, {Lefloch}, {Ceccarelli},
  {Steinacker}, {Castets}, \& {Loinard}}]{Bacm03}
{Bacmann}, A., {Lefloch}, B., {Ceccarelli}, C., {Steinacker}, J., {Castets},
  A., \& {Loinard}, L. 2003, \apjl, 585, L55

\bibitem[{{Bayet} {et~al.}(2009){Bayet}, {Aladro}, {Mart{\'{\i}}n}, {Viti}, \&
  {Mart{\'{\i}}n-Pintado}}]{Baye09c}
{Bayet}, E., {Aladro}, R., {Mart{\'{\i}}n}, S., {Viti}, S., \&
  {Mart{\'{\i}}n-Pintado}, J. 2009, \apj, 707, 126

\bibitem[{{Bayet} {et~al.}(2010{\natexlab{a}}){Bayet}, {Hartquist}, {Viti},
  {Williams}, \& {Bell}}]{Baye10a}
{Bayet}, E., {Hartquist}, T.~W., {Viti}, S., {Williams}, D.~A., \& {Bell},
  T.~A. 2010{\natexlab{a}}, ArXiv e-prints 2010arXiv1007.1533B

\bibitem[{{Bayet} {et~al.}(2008){Bayet}, {Viti}, {Williams}, \&
  {Rawlings}}]{Baye08a}
{Bayet}, E., {Viti}, S., {Williams}, D.~A., \& {Rawlings}, J.~M.~C. 2008, \apj,
  676, 978

\bibitem[{{Bayet} {et~al.}(2010{\natexlab{b}}){Bayet}, {Yates}, \&
  {Viti}}]{Baye10b}
{Bayet}, E., {Yates}, J., \& {Viti}, S. 2010{\natexlab{b}}, \mnras, submitted

\bibitem[{{Carilli} {et~al.}(2002){Carilli}, {Kohno}, {Kawabe}, {Ohta},
  {Henkel}, {Menten}, {Yun}, {Petric}, \& {Tutui}}]{Cari02}
{Carilli}, C.~L., {Kohno}, K., {Kawabe}, R., {Ohta}, K., {Henkel}, C.,
  {Menten}, K.~M., {Yun}, M.~S., {Petric}, A., \& {Tutui}, Y. 2002, \aj, 123,
  1838

\bibitem[{{Caselli} {et~al.}(2003){Caselli}, {van der Tak}, {Ceccarelli}, \&
  {Bacmann}}]{Case03}
{Caselli}, P., {van der Tak}, F.~F.~S., {Ceccarelli}, C., \& {Bacmann}, A.
  2003, \aap, 403, L37

\bibitem[{{Cazaux} {et~al.}(2010){Cazaux}, {Cobut}, {Marseille}, {Spaans}, \&
  {Caselli}}]{Caza10}
{Cazaux}, S., {Cobut}, V., {Marseille}, M., {Spaans}, M., \& {Caselli}, P.
  2010, ArXiv e-prints 2010arXiv1007.1061C

\bibitem[{{Ceccarelli} {et~al.}(1998){Ceccarelli}, {Castets}, {Loinard},
  {Caux}, \& {Tielens}}]{Cecc98}
{Ceccarelli}, C., {Castets}, A., {Loinard}, L., {Caux}, E., \& {Tielens},
  A.~G.~G.~M. 1998, \aap, 338, L43

\bibitem[{{Garrod} \& {Herbst}(2006)}]{Garr06}
{Garrod}, R.~T. \& {Herbst}, E. 2006, \aap, 457, 927

\bibitem[{{Gerin} {et~al.}(1992){Gerin}, {Combes}, {Wlodarczak}, {Jacq},
  {Guelin}, {Encrenaz}, \& {Laurent}}]{Geri92}
{Gerin}, M., {Combes}, F., {Wlodarczak}, G., {Jacq}, T., {Guelin}, M.,
  {Encrenaz}, P., \& {Laurent}, C. 1992, \aap, 259, L35

\bibitem[{{Israel} \& {Baas}(2003)}]{Isra03}
{Israel}, F.~P. \& {Baas}, F. 2003, \aap, 404, 495

\bibitem[{{Jacq} {et~al.}(1990){Jacq}, {Walmsley}, {Henkel}, {Baudry},
  {Mauersberger}, \& {Jewell}}]{Jacq90}
{Jacq}, T., {Walmsley}, C.~M., {Henkel}, C., {Baudry}, A., {Mauersberger}, R.,
  \& {Jewell}, P.~R. 1990, \aap, 228, 447

\bibitem[{{Jacq} {et~al.}(1993){Jacq}, {Walmsley}, {Mauersberger}, {Anderson},
  {Herbst}, \& {De Lucia}}]{Jacq93}
{Jacq}, T., {Walmsley}, C.~M., {Mauersberger}, R., {Anderson}, T., {Herbst},
  E., \& {De Lucia}, F.~C. 1993, \aap, 271, 276

\bibitem[{{Jiang} {et~al.}(2007){Jiang}, {Fan}, {Vestergaard}, {Kurk},
  {Walter}, {Kelly}, \& {Strauss}}]{Jian07}
{Jiang}, L., {Fan}, X., {Vestergaard}, M., {Kurk}, J.~D., {Walter}, F.,
  {Kelly}, B.~C., \& {Strauss}, M.~A. 2007, \aj, 134, 1150

\bibitem[{{Juarez} {et~al.}(2009){Juarez}, {Maiolino}, {Mujica}, {Pedani},
  {Marinoni}, {Nagao}, {Marconi}, \& {Oliva}}]{Juar09}
{Juarez}, Y., {Maiolino}, R., {Mujica}, R., {Pedani}, M., {Marinoni}, S.,
  {Nagao}, T., {Marconi}, A., \& {Oliva}, E. 2009, \aap, 494, L25

\bibitem[{{Knauth} {et~al.}(2003){Knauth}, {Andersson}, {McCandliss}, \&
  {Moos}}]{Knau03}
{Knauth}, D.~C., {Andersson}, B.-G., {McCandliss}, S.~R., \& {Moos}, H.~W.
  2003, \apjl, 596, L51

\bibitem[{{Le Teuff} {et~al.}(2000){Le Teuff}, {Millar}, \&
  {Markwick}}]{LeTe00}
{Le Teuff}, Y.~H., {Millar}, T.~J., \& {Markwick}, A.~J. 2000, \aaps, 146, 157

\bibitem[{{Leroy} {et~al.}(2006){Leroy}, {Bolatto}, {Walter}, \&
  {Blitz}}]{Lero06}
{Leroy}, A., {Bolatto}, A., {Walter}, F., \& {Blitz}, L. 2006, \apj, 643, 825

\bibitem[{{Lintott} {et~al.}(2005){Lintott}, {Viti}, {Williams}, {Rawlings}, \&
  {Ferreras}}]{Lint05}
{Lintott}, C.~J., {Viti}, S., {Williams}, D.~A., {Rawlings}, J.~M.~C., \&
  {Ferreras}, I. 2005, \mnras, 360, 1527

\bibitem[{{Loren} \& {Wootten}(1985)}]{Lore85}
{Loren}, R.~B. \& {Wootten}, A. 1985, \apj, 299, 947

\bibitem[{{Madden} {et~al.}(1997){Madden}, {Poglitsch}, {Geis}, {Stacey}, \&
  {Townes}}]{Madd97}
{Madden}, S.~C., {Poglitsch}, A., {Geis}, N., {Stacey}, G.~J., \& {Townes},
  C.~H. 1997, \apj, 483, 200

\bibitem[{{Mart{\'{\i}}n} {et~al.}(2006){Mart{\'{\i}}n}, {Mauersberger},
  {Mart{\'{\i}}n-Pintado}, {Henkel}, \& {Garc{\'{\i}}a-Burillo}}]{Mart06a}
{Mart{\'{\i}}n}, S., {Mauersberger}, R., {Mart{\'{\i}}n-Pintado}, J., {Henkel},
  C., \& {Garc{\'{\i}}a-Burillo}, S. 2006, \apjs, 164, 450

\bibitem[{{Mauersberger} {et~al.}(1995){Mauersberger}, {Henkel}, \&
  {Chin}}]{Maue95}
{Mauersberger}, R., {Henkel}, C., \& {Chin}, Y. 1995, \aap, 294, 23

\bibitem[{{Mauersberger} {et~al.}(1988){Mauersberger}, {Henkel}, {Jacq}, \&
  {Walmsley}}]{Maue88}
{Mauersberger}, R., {Henkel}, C., {Jacq}, T., \& {Walmsley}, C.~M. 1988, \aap,
  194, L1

\bibitem[{{Meijerink} \& {Spaans}(2005)}]{Meij05b}
{Meijerink}, R. \& {Spaans}, M. 2005, \aap, 436, 397

\bibitem[{{Meijerink} {et~al.}(2006){Meijerink}, {Spaans}, \&
  {Israel}}]{Meij06}
{Meijerink}, R., {Spaans}, M., \& {Israel}, F.~P. 2006, \apjl, 650, L103

\bibitem[{{Meyer} {et~al.}(1998){Meyer}, {Jura}, \& {Cardelli}}]{Meye98}
{Meyer}, D.~M., {Jura}, M., \& {Cardelli}, J.~A. 1998, \apj, 493, 222

\bibitem[{{Miettinen} {et~al.}(2009){Miettinen}, {Harju}, {Haikala},
  {Kainulainen}, \& {Johansson}}]{Miett09}
{Miettinen}, O., {Harju}, J., {Haikala}, L.~K., {Kainulainen}, J., \&
  {Johansson}, L.~E.~B. 2009, \aap, 500, 845

\bibitem[{{Millar} {et~al.}(1997){Millar}, {MacDonald}, \& {Gibb}}]{Mill97}
{Millar}, T.~J., {MacDonald}, G.~H., \& {Gibb}, A.~G. 1997, \aap, 325, 1163

\bibitem[{{Oliveira}(2003)}]{Oliv03}
{Oliveira}, C.~M. 2003, in Bulletin of the American Astronomical Society,
  Vol.~35, Bulletin of the American Astronomical Society, 1317--+

\bibitem[{{Parise} {et~al.}(2004){Parise}, {Castets}, {Herbst}, {Caux},
  {Ceccarelli}, {Mukhopadhyay}, \& {Tielens}}]{Pari04}
{Parise}, B., {Castets}, A., {Herbst}, E., {Caux}, E., {Ceccarelli}, C.,
  {Mukhopadhyay}, I., \& {Tielens}, A.~G.~G.~M. 2004, \aap, 416, 159

\bibitem[{{Parise} {et~al.}(2002){Parise}, {Ceccarelli}, {Tielens}, {Herbst},
  {Lefloch}, {Caux}, {Castets}, {Mukhopadhyay}, {Pagani}, \&
  {Loinard}}]{Pari02}
{Parise}, B., {Ceccarelli}, C., {Tielens}, A.~G.~G.~M., {Herbst}, E.,
  {Lefloch}, B., {Caux}, E., {Castets}, A., {Mukhopadhyay}, I., {Pagani}, L.,
  \& {Loinard}, L. 2002, \aap, 393, L49

\bibitem[{{Parra} {et~al.}(2007){Parra}, {Conway}, {Diamond}, {Thrall},
  {Lonsdale}, {Lonsdale}, \& {Smith}}]{Parr07}
{Parra}, R., {Conway}, J.~E., {Diamond}, P.~J., {Thrall}, H., {Lonsdale},
  C.~J., {Lonsdale}, C.~J., \& {Smith}, H.~E. 2007, \apj, 659, 314

\bibitem[{{Rawlings} {et~al.}(1992){Rawlings}, {Hartquist}, {Menten}, \&
  {Williams}}]{Rawl92}
{Rawlings}, J.~M.~C., {Hartquist}, T.~W., {Menten}, K.~M., \& {Williams}, D.~A.
  1992, \mnras, 255, 471

\bibitem[{{Riechers} {et~al.}(2009){Riechers}, {Walter}, {Carilli}, \&
  {Lewis}}]{Riec09}
{Riechers}, D.~A., {Walter}, F., {Carilli}, C.~L., \& {Lewis}, G.~F. 2009,
  \apj, 690, 463

\bibitem[{{Roberts} {et~al.}(2003){Roberts}, {Herbst}, \& {Millar}}]{Robe03}
{Roberts}, H., {Herbst}, E., \& {Millar}, T.~J. 2003, \apjl, 591, L41

\bibitem[{{Roberts} {et~al.}(2004){Roberts}, {Herbst}, \& {Millar}}]{Robe04}
---. 2004, \aap, 424, 905

\bibitem[{{Roberts} \& {Millar}(2000{\natexlab{a}})}]{Robe00a}
{Roberts}, H. \& {Millar}, T.~J. 2000{\natexlab{a}}, \aap, 364, 780

\bibitem[{{Roberts} \& {Millar}(2000{\natexlab{b}})}]{Robe00b}
---. 2000{\natexlab{b}}, \aap, 361, 388

\bibitem[{{R{\"o}llig} {et~al.}(2007){R{\"o}llig}, {Abel}, {Bell}, {Bensch},
  {Black}, {Ferland}, {Jonkheid}, {Kamp}, {Kaufman}, {Le Bourlot}, {Le Petit},
  \& {etc, }}]{Roel07}
{R{\"o}llig}, M., {Abel}, N.~P., {Bell}, T., {Bensch}, F., {Black}, J.,
  {Ferland}, G.~J., {Jonkheid}, B., {Kamp}, I., {Kaufman}, M.~J., {Le Bourlot},
  J., {Le Petit}, F., \& {etc, }. 2007, \aap, 467, 187

\bibitem[{{Roueff} \& {Gerin}(2003)}]{Roue03}
{Roueff}, E. \& {Gerin}, M. 2003, Space Science Reviews, 106, 61

\bibitem[{{Roueff} {et~al.}(2005){Roueff}, {Lis}, {van der Tak}, {Gerin}, \&
  {Goldsmith}}]{Roue05}
{Roueff}, E., {Lis}, D.~C., {van der Tak}, F.~F.~S., {Gerin}, M., \&
  {Goldsmith}, P.~F. 2005, \aap, 438, 585

\bibitem[{{Roueff} {et~al.}(2000){Roueff}, {Tin{\'e}}, {Coudert}, {Pineau des
  For{\^e}ts}, {Falgarone}, \& {Gerin}}]{Roue00}
{Roueff}, E., {Tin{\'e}}, S., {Coudert}, L.~H., {Pineau des For{\^e}ts}, G.,
  {Falgarone}, E., \& {Gerin}, M. 2000, \aap, 354, L63

\bibitem[{{Schilke} {et~al.}(1992){Schilke}, {Walmsley}, {Pineau Des Forets},
  {Roueff}, {Flower}, \& {Guilloteau}}]{Schi92}
{Schilke}, P., {Walmsley}, C.~M., {Pineau Des Forets}, G., {Roueff}, E.,
  {Flower}, D.~R., \& {Guilloteau}, S. 1992, \aap, 256, 595

\bibitem[{{Sembach} \& {Savage}(1996)}]{Sava96}
{Sembach}, K.~R. \& {Savage}, B.~D. 1996, \apj, 457, 211

\bibitem[{{Seymour}(2009)}]{Seym09}
{Seymour}, N. 2009, in Panoramic Radio Astronomy: Wide-field 1-2 GHz Research
  on Galaxy Evolution

\bibitem[{{Snow} {et~al.}(2002){Snow}, {Rachford}, \& {Figoski}}]{Snow02}
{Snow}, T.~P., {Rachford}, B.~L., \& {Figoski}, L. 2002, \apj, 573, 662

\bibitem[{{Sofia} {et~al.}(1997){Sofia}, {Cardelli}, {Guerin}, \&
  {Meyer}}]{Sofi97}
{Sofia}, U.~J., {Cardelli}, J.~A., {Guerin}, K.~P., \& {Meyer}, D.~M. 1997,
  \apjl, 482, L105+

\bibitem[{{Stark} {et~al.}(1999){Stark}, {van der Tak}, \& {van
  Dishoeck}}]{Star99}
{Stark}, R., {van der Tak}, F.~F.~S., \& {van Dishoeck}, E.~F. 1999, \apjl,
  521, L67

\bibitem[{{Tielens}(1983)}]{Tiel83}
{Tielens}, A.~G.~G.~M. 1983, \aap, 119, 177

\bibitem[{{van der Tak} {et~al.}(2007){van der Tak}, {Black}, {Sch{\"o}ier},
  {Jansen}, \& {van Dishoeck}}]{VanderTak07}
{van der Tak}, F.~F.~S., {Black}, J.~H., {Sch{\"o}ier}, F.~L., {Jansen}, D.~J.,
  \& {van Dishoeck}, E.~F. 2007, \aap, 468, 627

\bibitem[{{van Dishoeck} {et~al.}(1995){van Dishoeck}, {Blake}, {Jansen}, \&
  {Groesbeck}}]{VanDis95}
{van Dishoeck}, E.~F., {Blake}, G.~A., {Jansen}, D.~J., \& {Groesbeck}, T.~D.
  1995, \apj, 447, 760

\bibitem[{{Vastel} {et~al.}(2004){Vastel}, {Phillips}, \& {Yoshida}}]{Vast04}
{Vastel}, C., {Phillips}, T.~G., \& {Yoshida}, H. 2004, \apjl, 606, L127

\bibitem[{{Viti} {et~al.}(2004){Viti}, {Collings}, {Dever}, {McCoustra}, \&
  {Williams}}]{Viti04}
{Viti}, S., {Collings}, M.~P., {Dever}, J.~W., {McCoustra}, M.~R.~S., \&
  {Williams}, D.~A. 2004, \mnras, 354, 1141

\bibitem[{{Viti} \& {Williams}(1999)}]{Viti99}
{Viti}, S. \& {Williams}, D.~A. 1999, \mnras, 305, 755

\bibitem[{{Vrtilek} {et~al.}(1985){Vrtilek}, {Gottlieb}, {Langer}, {Thaddeus},
  \& {Wilson}}]{Vrti85}
{Vrtilek}, J.~M., {Gottlieb}, C.~A., {Langer}, W.~D., {Thaddeus}, P., \&
  {Wilson}, R.~W. 1985, \apjl, 296, L35

\bibitem[{{Walmsley} {et~al.}(1987){Walmsley}, {Hermsen}, {Henkel},
  {Mauersberger}, \& {Wilson}}]{Walm87}
{Walmsley}, C.~M., {Hermsen}, W., {Henkel}, C., {Mauersberger}, R., \&
  {Wilson}, T.~L. 1987, \aap, 172, 311

\bibitem[{{Watson}(1980)}]{Wats80}
{Watson}, W.~D. 1980, in Les Spectres des Mol{\'e}cules Simples au Laboratoire
  et en Astrophysique, 526--544

\bibitem[{{Wei{\ss}} {et~al.}(2007){Wei{\ss}}, {Downes}, {Neri}, {Walter},
  {Henkel}, {Wilner}, {Wagg}, \& {Wiklind}}]{Weis07}
{Wei{\ss}}, A., {Downes}, D., {Neri}, R., {Walter}, F., {Henkel}, C., {Wilner},
  D.~J., {Wagg}, J., \& {Wiklind}, T. 2007, \aap, 467, 955

\bibitem[{{White} \& {Frenk}(1991)}]{Whit91}
{White}, S.~D.~M. \& {Frenk}, C.~S. 1991, \apj, 379, 52

\bibitem[{{Woodall} {et~al.}(2007){Woodall}, {Ag{\'u}ndez}, {Markwick-Kemper},
  \& {Millar}}]{Wood07}
{Woodall}, J., {Ag{\'u}ndez}, M., {Markwick-Kemper}, A.~J., \& {Millar}, T.~J.
  2007, \aap, 466, 1197

\bibitem[{{Yin} {et~al.}(2010){Yin}, {Magrini}, {Matteucci}, {Lanfranchi},
  {Gon{\c c}alves}, \& {Costa}}]{Yin10}
{Yin}, J., {Magrini}, L., {Matteucci}, F., {Lanfranchi}, G.~A., {Gon{\c
  c}alves}, D.~R., \& {Costa}, R.~D.~D. 2010, ArXiv e-prints
  2010arXiv1005.3500Y

\bibitem[{{Zaritsky} {et~al.}(1994){Zaritsky}, {Kennicutt}, \&
  {Huchra}}]{Zari94}
{Zaritsky}, D., {Kennicutt}, R.~C., \& {Huchra}, J.~P. 1994, \apj, 420, 87

\end{thebibliography}

\begin{table*}
    \caption{Standard model parameters (See Model NS$_{3}$).}\label{tab:1}
    \begin{center}
    \begin{tabular}{c c c c c}
    \hline
    Parameter & Symbol & Typical Milky\\
    & & Way values\\
    \hline
    Collapse Mode$^{a}$ & $B$ & 0.1\\
    Initial number density (phase 1)& $n_{i}$ & 300 H cm$^{-3}$\\
    Final number density (phase 1 and 2)& $n_{f}$ & 1$\times 10^{7}$ cm$^{-3}$\\
    Temperature (phase 1) & $T_{1}$ & 10 K \\
    Temperature (phase 2) & $T_{2}$ & 300 K\\
    External UV radiation intensity & $I$ & 1 Habing \\
    Cosmic ray ionization rate & $\zeta$ & 1.3$\times 10^{-17}$
    s$^{-1}$\\
    Visual extinction & $A_{v}$ & 580.6 mag\\
    Gas:dust ratio & $d$ & 100 \\
    H$_{2}$ formation rate coefficient & $R$ & 1.0$\times 10^{-17}\times \sqrt{T}$ cm$^{3}$s$^{-1}$\\
    Metallicity & z$_{\odot}$ & solar values, see Table ~\ref{tab:2}\\
    \hline
    \end{tabular}
    \end{center}
    $^{a}$: The collapse is treated as a `modified
free-fall', as defined by \citet{Rawl92}, and the parameter $B$ is
introduced to allow for collapse in free-fall $B=$1.0, or somewhat
more slowly ($B<$1.0) if gas or magnetic pressure resists the
collapse.
\end{table*}

\begin{table}
    \caption{Initial elemental abundance ratios with respect to the total
    H-nuclei number, used as the standard values labelled (``ST'') in Table
    \ref{tab:2}. The initial elemental abundance ratios correspond here to a
    compilation of values from \citet{Sava96, Sofi97, Meye98, Snow02, Knau03}. }\label{tab:3}
    \begin{center}
    \begin{tabular}{c c}
    \hline
    & ST \\
    \hline
    C/H & 1.4$\times$10$^{-4}$ \\
    S/H & 1.4$\times$10$^{-6}$\\
    O/H & 3.2$\times$10$^{-4}$ \\
    N/H & 6.5$\times$10$^{-5}$ \\
    He/H & 7.5$\times$10$^{-2}$ \\
    Mg/H & 5.1$\times$10$^{-6}$ \\
    \hline
    \end{tabular}
    \end{center}
\end{table}

\begin{table*}
    \caption{Input parameters for the UCL\_Chem models (see
    Sect. \ref{sec:model}).
    The abbreviation ``ST'' represents the standard values listed in
    Table ~\ref{tab:1} and Table ~\ref{tab:3}. The abbreviation
    ``CB-A, -B, -C, -D, -E'' correspond to the values listed in Table ~\ref{tab:4}.}\label{tab:2}
    \begin{center}
    \begin{tabular}{c c c c c c c c c c}
    \hline
    Model & $n_{f}$ & Size & Metallicity$^{a}$ & Gas-to-dust & Ini. Elem. & $\zeta ^{b}$ & Temp.  &
    $I$ & Comb.\\
    & (cm$^{-3}$) & (pc) & (z$_{\odot}$) & mass ratio & Abund. ratios & ($\zeta_{\odot}$) & $T_{2}$ (K) &
    ($I_{\odot}$) & parameters\\
    \hline
    Normal spiral&&&&&&&&\\
    NS$_{1}$ & 10$^{5}$ & 1.00  & 1 & 100 & ST & 1 & 300 & 1 & CB-A\\
    NS$_{2}$ & 10$^{6}$ & 0.15 & 1 & 100 & ST & 1 & 300 & 1 & CB-B\\
    NS$_{3}$ & 10$^{7}$ & 0.03 & 1 & 100 & ST & 1 & 300 & 1 & ST\\
    \hline
    Starburst&&&&&&&&\\
    SB$_{1}$ & 10$^{5}$ & 1.00 & 1 & 100 & ST & 1 & 500 & $10^{3}$ & CB-C\\
    SB$_{2}$ & 10$^{6}$ & 0.15 & 1 & 100 & ST & 1 & 500 & $10^{3}$ & CB-D\\
    SB$_{3}$ & 10$^{7}$ & 0.03 & 1 & 100 & ST & 1 & 500 & $10^{3}$ & CB-E\\
    \hline
    Cosmic-Rays &&&&&&&&\\
    Enhanced&&&&&&&&\\
    SB$_{1}+$ & 10$^{5}$ & 1.00 & 1 & 100 & ST & 100 & 300 & 1 & CB-A\\
    SB$_{2}+$ & 10$^{6}$ & 0.15 & 1 & 100 & ST & 100 & 300 & 1 & CB-B\\
    SB$_{3}+$ & 10$^{7}$ & 0.03 & 1 & 100 & ST & 100 & 300 & 1 & ST\\
    \hline
    Low metallicity&&&&&&&&\\
    Low-met$_{1}$ & 10$^{5}$ & 1.00 & 1/5 & 500 & ST/5 & 1 & 500 & $10^{3}$ & CB-F\\
    Low-met$_{2}$ & 10$^{6}$ & 0.15 & 1/5 & 500 & ST/5 & 1 & 500 & $10^{3}$ & CB-G\\
    Low-met$_{3}$ & 10$^{7}$ & 0.03 & 1/5 & 500 & ST/5 & 1 & 500 & $10^{3}$ & CB-H\\
    \hline
    High redshift&&&&&&&&\\
    High-z$_{1}$ & 10$^{5}$ & 1.00 & 1/5 & 500 & ST/5 & 100 & 500 & 10$^{3}$ & CB-F\\
    High-z$_{2}$ & 10$^{6}$ & 0.15 & 1/5 & 500 & ST/5 & 100 & 500 & 10$^{3}$ & CB-G\\
    High-z$_{3}$ & 10$^{7}$ & 0.03 & 1/5 & 500 & ST/5 & 100 & 500 & 10$^{3}$ & CB-H\\
    \hline
    \end{tabular}
    \end{center}
    $^{a}$ : z$_{\odot}$ = 1 corresponds to solar values of the elemental
    abundances ratios; $^{b}$ : expressed in units of
    $\zeta_{\odot}$ = 1.3$\times 10^{-17}$ s$^{-1}$.
\end{table*}

\begin{landscape}
\begin{table}
    \caption{Combinations of parameters coupled with the metallicity and used
    in
    Table. ~\ref{tab:2} (see Sect. ~\ref{sec:model}).}\label{tab:4}
    \hspace*{-0.8cm}
    \begin{tabular}{c c c c c c c c c c}
    \hline
    & ST & CB-A & CB-B & CB-C & CB-D & CB-E & CB-F & CB-G & CB-H\\
    \hline
    $A_{v}$ & 580.6 & 194.9 &  291.3 & 194.9 & 291.3 & 580.6 & 40.6 & 59.9 & 117.7\\
    (mag) &&&&&&&&&\\
    \hline
     ratio of the number &&&&&&&&&\\
     densities of grains & 1.0$\times$10$^{-12}$ & 1.0$\times$10$^{-12}$ & 1.0$\times$10$^{-12}$
     & 1.0$\times$10$^{-12}$ & 1.0$\times$10$^{-12}$ & 1.0$\times$10$^{-12}$
     & 2.0$\times$10$^{-13}$ & 2.0$\times$10$^{-13}$ & 2.0$\times$10$^{-13}$\\
    to hydrogen nuclei&&&&&&&&&\\
    \hline
    H$_{2}$ form.$^{a}$ & 1.7$\times$10$^{-16}$ & 1.7$\times$10$^{-16}$ & 1.7$\times$10$^{-16}$
    & 2.2$\times$10$^{-16}$ & 2.2$\times$10$^{-16}$ & 2.2$\times$10$^{-16}$
    & 4.5$\times$10$^{-17}$ & 4.5$\times$10$^{-17}$ & 4.5$\times$10$^{-17}$\\
    rate coeff. &&&&&&&&&\\
    \hline
    \end{tabular}

    $^{a}$ : The H$_{2}$ formation rate coefficient is computed at the beginning
    of the phase 2 of the UCL hot core model and it is
    expressed in cm$^{3}$s$^{-1}$.
\end{table}
\end{landscape}

\begin{landscape}
\begin{table}
\caption{Deuterated detectable in extragalactic star forming
regions and their corresponding D/H ratios (at a time=10$^{6}$
yrs) for five models representative of five types of galaxies (see
Sect. \ref{sec:model}). The limit of detectability has been taken
to be [n(X)/n$_{H}$] = $1\times 10^{-12}$, as is typical for dense
gas in the Milky Way. Below this limit, we assumed the species as
not detectable (symbol ``-''). Otherwise, they are marked with the
symbol ``+''. When the fractional abundance of a deuterated
species is above [n(X)/n$_{H}$] = $1\times 10^{-10}$, the symbol
``++'' is used.}\label{tab:5}
  \hspace{-1.1cm}
  \resizebox{24cm}{!}{
  \begin{tabular}{|l|c|c|c||c|c|c||c|c|c||c|c|c||c|c|c|}
\hline &\multicolumn{3}{c}{Normal Spiral}
&\multicolumn{3}{c}{Starburst} &\multicolumn{3}{c}{Cosmic-
Rays} &\multicolumn{3}{c}{Low}
&\multicolumn{3}{c}{High}\\
& & & & & & & & Enhanced & & & metallicity & & & redshift&\\
\cline{2-16}
 & NS$_{1}$&NS$_{2}$&NS$_{3}$&SB$_{1}$&SB$_{2}$&SB$_{3}$&SB$_{1}+$&SB$_{2}+$&
SB$_{3}+$&Low-met$_{1}$&Low-met$_{2}$&Low-met$_{3}$&High-z$_{1}$&
High-z$_{2}$&High-z$_{3}$\\
\hline
HDCO & ++ & ++ & ++ & ++ & ++ & ++ & + & + & + & + & + & + & + & + & ++ \\
D$_{2}$CO & + & + & + & + & + & + & - & - & - & - & - & - & - & - & - \\
DCN & ++ & + & + & ++ & + & + & ++ & ++ & ++ & ++ & ++ & + & ++ & ++ & ++ \\
DNC & + & + & - & - & - & - & - & - & - & - & - & - & - & - & - \\
DC$_{3}$N & - & - & - & + & - & - & - & - & - & - & - & - & - & - & - \\
DCO$^{+}$ & - & - & - & - & - & - & + & - & - & - & - & - & - & - & - \\
H$_{2}$D$^{+}$ & - & - & - & - & - & - & - & - & - & - & - & - & - & - & - \\
HDO & ++ & ++ & ++ & ++ & ++ & ++ & ++ & ++ & ++ & ++ & ++ & ++ & ++ & ++ & ++ \\
C$_{2}$D & ++ & ++ & ++ & ++ & ++ & ++ & - & - & - & ++ & ++ & ++ & ++ & ++ & ++ \\
HDCS & ++ & ++ & ++ & ++ & ++ & ++ & - & - & - & ++ & ++ & ++ & + & + & + \\
HDS & + & + & + & - & - & + & - & - & + & - & - & - & - & - & - \\
NH$_{2}$D & ++ & ++ & ++ & ++ & ++ & ++ & + & - & - & + & + & - & - & - & - \\
N$_{2}$D$^{+}$ & - & - & - & - & - & - & - & - & + & - & - & - & - & - & - \\
CH$_{3}$OD & + & + & + & + & + & + & - & - & - & - & - & - & - & - & - \\
CH$_{2}$DOH & + & + & + & + & + & + & - & - & - & + & - & - & - & - & - \\
CH$_{2}$DCN & - & - & - & - & - & - & - & - & - & - & - & - & - & - & - \\
\hline \hline
D$_{2}$CO/H$_{2}$CO  & $5.1 \times 10^{-6}$ & $8.8 \times 10^{-7}$& $6.2 \times 10^{-6}$& $2.8 \times 10^{-5}$ & $1.6 \times 10^{-6}$&$4.0 \times 10^{-5}$ &$1.2 \times 10^{-4}$&$4.2 \times 10^{-4}$&$2.7 \times 10^{-5}$&$3.1 \times 10^{-6}$&$8.8 \times 10^{-6}$&$3.7 \times 10^{-6}$&$5.5 \times 10^{-7}$&$3.8 \times 10^{-6}$&$2.6 \times 10^{-6}$\\
DCN/HCN & $1.4 \times 10^{-4}$ &$7.2 \times 10^{-5}$&$2.0 \times 10^{-4}$&$3.1 \times 10^{-4}$&$7.6 \times 10^{-4}$&$4.2 \times 10^{-3}$&$1.6 \times 10^{-3}$&$8.7 \times 10^{-3}$&$2.0 \times 10^{-3}$&$6.0 \times 10^{-4}$&$3.5 \times 10^{-3}$&$3.1 \times 10^{-3}$&$6.3 \times 10^{-4}$&$2.6 \times 10^{-3}$&$8.2 \times 10^{-3}$\\
DNC/HNC & $4.8 \times 10^{-5}$&$2.4 \times 10^{-5}$&$2.8 \times 10^{-5}$&$6.5 \times 10^{-6}$&$6.5 \times 10^{-6}$&$2.2 \times 10^{-5}$&$3.3 \times 10^{-10}$&$2.3 \times 10^{-11}$&$5.4 \times 10^{-11}$&$1.9 \times 10^{-7}$&$5.9 \times 10^{-8}$&$8.8 \times 10^{-8}$&$9.6 \times 10^{-9}$&$3.2 \times 10^{-9}$&$2.6 \times 10^{-10}$\\
NH$_{2}$D/NH$_{3}$ & $2.7\times 10^{-5}$&$1.3 \times 10^{-5}$&$1.4 \times 10^{-5}$&$3.3 \times 10^{-5}$&$2.4 \times 10^{-5}$&$2.6 \times 10^{-5}$&$5.0 \times 10^{-5}$&$1.5 \times 10^{-5}$&$3.0 \times 10^{-5}$&$1.3 \times 10^{-6}$&$1.5 \times 10^{-6}$&$1.8 \times 10^{-6}$&$1.8 \times 10^{-6}$&$1.6 \times 10^{-6}$&$2.1 \times 10^{-6}$\\
DCO$^{+}$/HCO$^{+}$ &$3.9 \times 10^{-4}$&$2.0 \times 10^{-4}$&$2.1 \times 10^{-4}$&$3.0 \times 10^{-4}$&$2.4 \times 10^{-4}$&$4.2 \times 10^{-4}$&$6.5 \times 10^{-3}$&$2.2 \times 10^{-3}$&$2.9 \times 10^{-4}$&$8.2 \times 10^{-5}$&$1.5 \times 10^{-4}$&$2.4 \times 10^{-4}$&$7.1 \times 10^{-4}$&$5.6 \times 10^{-4}$&$5.4 \times 10^{-4}$\\
HDCO/H$_{2}$CO &$1.0 \times 10^{-3}$&$5.7 \times 10^{-4}$&$5.0 \times 10^{-4}$&$3.3 \times 10^{-3}$&$9.2 \times 10^{-4}$&$2.5 \times 10^{-4}$&$1.7 \times 10^{-2}$&$3.3 \times 10^{-2}$&$2.5 \times 10^{-1}$&$6.2 \times 10^{-3}$&$9.5 \times 10^{-3}$&$1.4 \times 10^{-3}$&$7.5 \times 10^{-3}$&$6.7 \times 10^{-2}$&$1.4 \times 10^{-1}$\\
CH$_{3}$OD/CH$_{3}$OH&$1.1 \times 10^{-3}$&$7.2 \times 10^{-4}$&$7.3 \times 10^{-4}$&$9.3 \times 10^{-4}$&$5.8 \times 10^{-4}$&$6.4 \times 10^{-4}$&$1.5 \times 10^{-2}$&$1.2 \times 10^{-2}$&$3.6 \times 10^{-3}$&$8.0 \times 10^{-3}$&$6.7 \times 10^{-3}$&$9.1 \times 10^{-3}$&$4.6 \times 10^{-3}$&$9.0 \times 10^{-3}$&$8.3 \times 10^{-2}$\\
CH$_{2}$DOH/CH$_{3}$OH  &$1.5 \times 10^{-3}$&$1.2 \times 10^{-3}$&$1.1 \times 10^{-3}$&$1.5 \times 10^{-3}$&$8.7 \times 10^{-4}$&$5.8 \times 10^{-4}$&$1.9 \times 10^{-2}$&$1.5 \times 10^{-2}$&$1.3 \times 10^{-2}$&$9.5 \times 10^{-2}$&$6.6 \times 10^{-2}$&$2.4 \times 10^{-2}$&$7.5 \times 10^{-3}$&$3.0 \times 10^{-2}$&$3.3 \times 10^{-1}$\\
CH$_{2}$DCN/CH$_{3}$CN  &$7.2 \times 10^{-4}$&$1.4 \times 10^{-3}$&$5.4 \times 10^{-4}$&$6.1 \times 10^{-3}$&$1.0 \times 10^{-3}$&$1.1 \times 10^{-4}$&$3.1 \times 10^{-5}$&$2.9 \times 10^{-5}$&$3.8 \times 10^{-5}$&$4.1 \times 10^{-3}$&$3.9 \times 10^{-3}$&$4.2 \times 10^{-4}$&$7.2 \times 10^{-3}$&$6.8 \times 10^{-2}$&$5.3 \times 10^{-2}$\\
H$_{2}$D$^{+}$/H$_{3}^{+}$  &$7.5 \times 10^{-5}$&$7.6 \times 10^{-5}$&$7.6 \times 10^{-5}$&$5.6 \times 10^{-5}$&$5.6 \times 10^{-5}$&$5.3 \times 10^{-6}$&$6.6 \times 10^{-5}$&$6.1 \times 10^{-5}$&$5.6 \times 10^{-5}$&$1.8 \times 10^{-5}$&$1.7 \times 10^{-5}$&$1.7 \times 10^{-5}$&$1.9 \times 10^{-5}$&$1.1 \times 10^{-5}$&$4.7 \times 10^{-6}$\\
N$_{2}$D$^{+}$/N$_{2}$H$^{+}$  &$5.2 \times 10^{-4}$&$2.1 \times 10^{-4}$&$2.4 \times 10^{-4}$&$5.3 \times 10^{-4}$&$5.4 \times 10^{-4}$&$1.4 \times 10^{-3}$&$1.2 \times 10^{-3}$&$9.4 \times 10^{-4}$&$1.9 \times 10^{-4}$&$2.9 \times 10^{-4}$&$4.8 \times 10^{-4}$&$8.1 \times 10^{-4}$&$7.1 \times 10^{-4}$&$1.1 \times 10^{-3}$&$1.5 \times 10^{-3}$\\
\hline
\end{tabular}}
\end{table}
\end{landscape}

\end{document}